\documentclass[aps,prl,amsmath,amssymb,twocolumn,preprintnumbers,nofootinbib,longbibliography,superscriptaddress,10pt]{revtex4-1}
\usepackage[utf8]{inputenc} 
\usepackage{graphicx}
\usepackage{url}
\usepackage[bookmarks, pagebackref=false]{hyperref}
\usepackage[usenames,dvipsnames]{xcolor}
\definecolor{orange}{cmyk}{0,0.5,1,0}
\definecolor{rossoCP3}{cmyk}{0,.88,.77,.40}
\definecolor{graa}{rgb}{0.8,0.8,0.8}
\definecolor{blaa}{rgb}{0.2,0.2,0.6}
\hypersetup{
  colorlinks,
  bookmarksopen,
  bookmarksnumbered,
  citecolor=blaa, 		
  linkcolor=rossoCP3,	
  urlcolor=rossoCP3,			
}
\usepackage{bm}
\usepackage{pxfonts}
\usepackage{amsmath}
\usepackage{amssymb}
\usepackage{amsfonts}
\usepackage{graphicx}
\usepackage{amssymb}
\usepackage{mathrsfs}
\usepackage{dsfont}%
\usepackage{slashed}
\usepackage{braket}
\usepackage{natbib}

\usepackage{adjustbox}
\usepackage{tikz}
\usetikzlibrary{patterns}
\usetikzlibrary{decorations.pathmorphing}
\usetikzlibrary{decorations.markings}
\usepgflibrary{arrows.meta}
\tikzstyle{box}=[pattern=north west lines, pattern color=colD, draw opacity=0.5,fill opacity=0.2,draw=colD,rounded corners=10]
\tikzstyle{op}=[circle, draw=black, fill=white,inner sep=0pt,minimum size=0pt]
\tikzstyle{prop}=[-, draw=colA, line width=1pt]
\tikzstyle{legOp}=[-, draw=colA, line width=1pt]
\tikzstyle{legExt}=[-, draw=colA, line width=1pt]

\definecolor{colA}{HTML}{c19277}
\definecolor{colB}{HTML}{e1bc91}
\definecolor{colD}{HTML}{62959c}

\newcommand{\beq}{\begin{eqnarray}}
\newcommand{\eeq}{\end{eqnarray}}
\newcommand{\NO}{~\nonumber \\}
\newcommand{\p}{{\cal P}\exp}

\newcommand{\bmp}{\noindent\begin{minipage}{16cm}}
\newcommand{\emp}{\end{minipage}\vskip 7mm} 

\newcommand   \f  {\phi}
\newcommand   \x   {\times}
\renewcommand \l  {\lambda}
\newcommand   \<   {\langle}
\renewcommand \>   {\rangle}

\newcommand*{\del}{\mathop{\mathrm{{}\partial}}\mathopen{}}

\renewcommand{\d}{\textrm{d}}

\def\lsim{\mathrel{\rlap{\lower4pt\hbox{\hskip1pt$\sim$}}
    \raise1pt\hbox{$<$}}}                
\def\gsim{\mathrel{\rlap{\lower4pt\hbox{\hskip1pt$\sim$}}
    \raise1pt\hbox{$>$}}}                

\def\k{\kappa}
\def\r{\rho}
\def\a{\alpha}

\def\g{\gamma}

\def\d{\delta}

\def\e{\epsilon}

\def\p{\pi}

\def\c{\chi}

\def\th{\theta}

\def\m{\mu}
\def\n{\nu}
\def\o{\omega}

\def\l{\lambda}
\def\L{\Lambda}
\def\s{\sigma}

\def\x{{\xi}}
\def\ve{{\varepsilon}}
\def\h{{\eta}}
\def\j{{\psi}}

\def\t{{\tau}}
\def\vth{{\vartheta}}

\def\f{{\phi}}

\def\vf{{\varphi}}

\def\vf{{\varphi}}

\def\pa{{\partial}}

\def\cs{{\cal S}}

\def\cd{{\cal D}}

\def\IR{\relax{\rm I\kern-.18em R}}

\def\hve{{\hat\ve}}

\def\IR{\relax{\rm I\kern-.18em R}}
\def\IL{\relax{\rm I\kern-.18em L}}

\def\inv{^{\raise.15ex\hbox{${\scriptscriptstyle -}$}\kern-.05em 1}}

\def\ca{{\cal A}}
\def\cn{{\cal N}}

\def\cj{{\cal J}}

\def\cw{{\cal W}}
\def\cf{{\cal F}}
\def\ct{{\cal T}}
\def\cz{{\cal Z}}
\def\co{{\cal O}}

\DeclareMathAlphabet{\dutchcal}{U}{dutchcal}{m}{n}

\def\){\right)}
\def\({\left( }
\def\]{\right] }
\def\[{\left[ }

\def\NO{\nonumber}

\newcommand{\be}{\begin{equation}}
\newcommand{\ee}{\end{equation}}

\def\bea{\begin{eqnarray}}
\def\eea{\end{eqnarray}}

\def\bal#1\eal{\begin{align}#1\end{align}}

\def\bald{\begin{aligned}}
\def\eald{\end{aligned}}

\def\bsub{\begin{subequations}}
\def\esub{\end{subequations}}

\def\beqx{\begin{displaymath}}
\def\eeqx{\end{displaymath}}

\newcommand{\bmat}{\left(\begin{array}}
\newcommand{\emat}{\end{array}\right)}

\def\a{\alpha}

\def\c{\chi}
\def\d{\delta}
\def\e{\epsilon}
\def\f{\phi}
\def\g{\gamma}
\def\h{\eta}
\def\j{\psi}
\def\k{\kappa}
\def\l{\lambda}
\def\m{\mu}
\def\n{\nu}
\def\o{\omega}
    
\def\p{\pi}

    \def\th{\theta}
\def\r{\rho}
\def\s{\sigma}
\def\t{\tau}
\def\x{\xi}

\def\F{\Phi}

\def\L{\Lambda}

\def\ve{\varepsilon}

    \def\vth{\vartheta}

\def\vf{\varphi}

\def\ca{{\cal A}}

\def\cd{{\cal D}}

\def\cf{{\cal F}}

\def\cj{{\cal J}}

\def\cl{{\cal L}}

\def\cn{{\cal N}}
\def\co{{\cal O}}

\def\cs{{\cal S}}
\def\ct{{\cal T}}

\def\cw{{\cal W}}

\def\cz{{\cal Z}}

\def\dc#1{\ensuremath{\dutchcal{#1}}} 

\def\bo{{\raise-.3ex\hbox{\large$\Box$}}}               
\def\pa{\partial}                                       
\def\face{{\raise.2ex\hbox{$\displaystyle \bigodot$}\mskip-2.2mu \llap {$\ddot
        \smile$}}}                                   
\def\>{\rangle}                                      
\def\<{\langle}                                      

\def\wt#1{\widetilde{#1}}                            
\def\Hat#1{\widehat{#1}}                             
\def\lbar#1{\ensuremath{\overline{#1}}}              
\def\leftrightarrowfill{$\mathsurround=0pt \mathord\leftarrow \mkern-6mu
        \cleaders\hbox{$\mkern-2mu \mathord- \mkern-2mu$}\hfill
        \mkern-6mu \mathord\rightarrow$}        
\def\dvec#1{\vbox{\ialign{##\crcr
        \leftrightarrowfill\crcr\noalign{\kern-1pt\nointerlineskip}
        $\hfil\displaystyle{#1}\hfil$\crcr}}}           

\def\-{\hphantom{-}}

\begin{document}

\preprint{KIAS-P26081}

\title{Interpolating between 4d $\cn=1$ SCFT supercurrent multiplet anomalies}

\author{Pantelis {\sc Panopoulos}}\email{ppanopoulos@kias.re.kr}
\affiliation{School of Physics, Korea Institute for Advanced Study, Seoul 02455, Korea}
\author{Ioannis {\sc Papadimitriou}}\email{ioannis.papadimitriou@phys.uoa.gr}
\affiliation{Division of Nuclear and Particle Physics, Department of Physics,
National and Kapodistrian University of Athens, 157 84 Athens, Greece}
\affiliation{Department of Physics, College of Science, Kyung Hee University, Seoul 02447, Korea}

\begin{abstract}
A renormalized SCFT can be coupled to a number of background off-shell supergravity multiplets, defining different current multiplets and superconformal anomalies. We focus on the relation between the conformal and Ferrara-Zumino multiplets of generic $\cn=1$ SCFTs in four dimensions, defined respectively by $\cn=1$ conformal and old-minimal supergravity. We show that the conformal supergravity partition function,  when lifted to a functional of the old-minimal background fields through a suitable Wess-Zumino action, is related to the old-minimal partition function by a local counterterm that we determine explicitly in components. This counterterm interpolates between two different old-minimal supergravity schemes with different superconformal anomalies. Part of the counterterm depends on the old-minimal auxiliary scalar and provides the explicit component form of the superspace counterterm constructed in \cite{Kuzenko:2019vvi}. However, the full counterterm that interpolates between the two multiplets contains additional terms, independent of the old-minimal auxiliary scalar, which are not present in this superspace counterterm. These additional terms play a key role in the comparison of physical observables between the two multiplets.
\end{abstract}
\maketitle
\small

\section{Introduction}

Placing a supersymmetric quantum field theory (SQFT) on a curved background is a useful technique for probing a number of different aspects of the theory. It facilitates the study of quantum anomalies and Casimir charges, leads to powerful computational methods such as supersymmetric localization, and arises naturally in the context of holography, to mention a few. Background fields for SQFTs can be turned on consistently by coupling the theory to a suitable off-shell supergravity \cite{Festuccia:2011ws}. However, there is often a choice of different off-shell supergravities a theory can be coupled to, resulting in distinct supercurrent multiplets \cite{Komargodski:2010rb}. When two such multiplets exist simultaneously, they may be related by improvement terms or local counterterms, but can still lead to apparently different results. Moreover, one or the other multiplet might be more natural in the context of specific applications. In the case of superconformal field theories (SCFTs), for example, the conformal current multiplet is implicitly selected in the context of holography, while supersymmetric localization picks a different current multiplet. In order to compare physical observables computed with different methods, it is therefore necessary to know how different current multiplets are related.  

Here, we focus on the relation between the conformal and Ferrara-Zumino multiplets of generic $\cn=1$ SCFTs in four dimensions, defined respectively by $\cn=1$ conformal and old-minimal supergravity. The conformal multiplet possesses a Q-supersymmetry anomaly \cite{Papadimitriou:2017kzw,An:2017ihs,Papadimitriou:2019gel,Katsianis:2019hhg,Imbimbo:2023sph,Imbimbo:2025ffw}. Its identification in \cite{Papadimitriou:2017kzw} resolved apparent tension in the comparison between supersymmetric observables computed using supersymmetric localization and holographic techniques \cite{Genolini:2016ecx}. Subsequently, it was shown in \cite{Kuzenko:2019vvi} and \cite{Katsianis:2020hzd} that, when conformal supergravity is {\em lifted} to old-minimal supergravity, there exists a local counterterm -- a local functional of the {\em old-minimal fields} -- that removes the Q-supersymmetry anomaly. The term "lifted" here is key; lifting requires a Wess-Zumino (WZ) term. There exists no local counterterm within conformal supergravity that removes the Q-supersymmetry anomaly -- it would violate the WZ consistency conditions of the conformal supergravity algebra \cite{Papadimitriou:2019gel}.

The local counterterm constructed in \cite{Kuzenko:2019vvi} relies on the superspace WZ anomaly effective action of $\cn=1$ supergravity obtained in \cite{Schwimmer:2010za}. This is equivalent to the WZ action of old-minimal supergravity in the presence of local super-Weyl symmetry, which we derive anew in components in the companion work \cite{suzywzactions}. We show that the component form of the superspace counterterm of \cite{Kuzenko:2019vvi} coincides with the part of our local counterterm that depends on the old-minimal auxiliary scalar, $\hat\cf$, given in \eqref{eq:OM-counterterm-F}. This does define a functional that depends only on the (lifted) conformal supergravity fields -- what is termed ``WZ gauge'' in \cite{Kuzenko:2019vvi} -- but this functional differs from the canonical functional of conformal supergravity by the additional, $\hat\cf$-independent counterterm given in \eqref{eq:OM-counterterm-0}. Some terms of this counterterm were determined in \cite{Katsianis:2020hzd} for specific values of the anomaly coefficients by studying a number of flat space current multiplet correlation functions in the free and massless WZ model. Here, we present the complete local counterterm that interpolates between the two current multiplets for arbitrary values of the anomaly coefficients $a$ and $c$. The expression for the $\hat\cf$-dependent part of the counterterm in \eqref{eq:OM-counterterm-F} is exact in the gravitino. The same we believe holds for the terms proportional to the $a$ anomaly coefficient in the $\hat\cf$-independent counterterm in \eqref{eq:OM-counterterm-0}, although we have confirmed the counterterm to quadratic order in the gravitino. The terms proportional to $a-c$ in the $\hat\cf$-independent counterterm in \eqref{eq:OM-counterterm-0} are given only to quadratic order in the gravitino.

\section{Conformal current multiplet and anomalies}
\label{sec:}

The conformal supercurrent multiplet is defined by coupling a renormalized SCFT to background off-shell conformal supergravity \cite{Festuccia:2011ws}. The field content of $\cn=1$ conformal supergravity consists of the vierbein $e_\m{}^a$, the graviphoton $A_\m$, and a Majorana gravitino $\j_{\m}$, comprising 5+3 bosonic and 8 fermionic off-shell degrees of freedom. Given the renormalized partition function of the theory on an arbitrary supergravity background,
\be\label{sugraW}
\cw[e,A,\psi]\equiv -i\log\cz[e,A,\psi]\,,
\ee
the conformal supercurrent multiplet operators are defined by
\be\label{consistent-currents}
\<\ct^\m{}_a\>=e^{-1}\frac{\d\mathcal W}{\d e_\m{}^a}\,,\quad \<\cj^\m\>=e^{-1}\frac{\d\mathcal W}{\d A_\m}\,,\quad \<\cs^\m\>=e^{-1}\frac{\d\mathcal W}{\d \lbar\psi_\m}\,,
\ee
where $e\equiv\det(e_\m{}^a)$ and the brackets $\<\cdot\>$ denote renormalized correlation functions on an arbitrary supergravity background.

$\cn=1$ conformal supergravity is obtained by gauging the superconformal algebra $\frak su$(2,2$|$1) \cite{Kaku:1977pa,Kaku:1977rk,Kaku:1978nz,Townsend:1979ki,VanNieuwenhuizen:1981ae,deWit:1981vgr,deWit:1983qkc,Fradkin:1985am,Freedman:2012zz}, as is summarized in table~\ref{gauging}. The fields $e_\m{}^a$, $A_\m$ and $\j_{\m}$ correspond to gauge connections for translations, R-symmetry and Q-supersymmetry transformations, respectively, while all other connections are either gauge-fixed to zero or expressed as composites of $e_\m{}^a$, $A_\m$ and $\j_{\m}$ through curvature constraints. Specifically, the dilatations connection, $b_\m$, is set to zero by gauge-fixing local special conformal transformations, while the local Lorentz, special conformal and S-supersymmetry connections, respectively $\o_\m{}^{ab}$, $f_\m{}^a$ and $\f_\m$, become composite by imposing suitable curvature constraints, namely\footnote{Here we have set the parameter $k$ in the more general expressions given in \cite{suzywzactions} to zero. This choice coincides with the convention of e.g. \cite{Kaku:1978nz, Fradkin:1985am}, but differs from that of \cite{Freedman:2012zz}.}
\bal\label{composites}
\o_\m{}^{ab}=&\;\o_\m{}^{ab}(e)+\tfrac14(\bar\j^a\g_\m\j^b+\bar\j_\m\g^a\j^b-\bar\j_\m\g^b\j^a)\,,\NO\\
\f_\m=&\; -\tfrac16(4\d^{[\r}_\m\d^{\s]}_\n+i\g^5 \e_{\m\n}{}^{\r\s})\g^\n \cd_\r\j_\s\,,\NO\\
f_\m{}^a=&\;-\tfrac14\big(\Hat{\mathscr{R}}_{\m}{}^{a}-\tfrac16 \Hat{\mathscr{R}} e_\m{}^a\big)-\tfrac{1}{8}\bar\j^\n\g_\m \mathscr{R}_{\n}{}^a(Q)-\tfrac{1}{6}\wt{\mathscr{R}}_\m{}^a(T)\,,
\eal
where $\wt{\mathscr{R}}_\m{}^a(T)=\tfrac12\e_{\m\n\r\s}e^{\n a}\mathscr{R}^{\r\s}(T)$ and
\bal\label{curvatures}
\Hat {\mathscr{R}}_{\m\n}{}^{ab}=&\;R_{\m\n}{}^{ab}(\o)-\bar\j_{[\m}\g^{ab}\f_{\n]}\,,\NO\\
\mathscr{R}_{\m\n}(Q)=&\;2\cd_{[\m}\j_{\n]}-2\g_{[\m}\f_{\n]}\,,\NO\\
\mathscr{R}_{\m\n}(T)=\,& F_{\m\n}-\tfrac{3i}{2}\bar\j_{[\m}\g^5\f_{\n]}\,.
\eal
\begin{table}
\begin{tabular}{c|ccccccc}
$T_A$ & $P_a$ & $M_{[ab]}$ & $D$ & $K_a$ &   $U(1)_R$ & $Q_\a$ &   $S_\a$\vspace{.1cm}\\
\hline\hline\rule{.0cm}{.5cm} 
$\vth^A$ & $\x^a$ & $\l^{ab}$ &  $\s$ & $\l_K^a$ &  $\th$ & $\ve^\a$ & $\h^\a$ \vspace{.1cm}\\  
\hline\rule{.0cm}{.5cm} 
$\mathscr{A}^A_\m$ & $ e_\m{}^a$ & $\o_\m{}^{ab}$ & {$b_\m$}  & {$f_\m{}^a$ }  & $A_\m$ & $\j^\a_\m$  &  {$\phi^\a_\m$} \vspace{.1cm}\\   
\hline\hline\rule{.0cm}{.5cm} 
$\ca_A$   &  $-$ & $-$ & $\ca_W$ & $-$ & $ \ca_R$ & $\ca_Q$ &$\ca_S$
\end{tabular}
\caption{List of the $\frak su$(2,2$|$1) generators $T_A$, the corresponding gauge parameters $\vth^A$, gauge connections $\mathscr{A}^A_\m$ and superconformal anomalies $\ca_A$. The connections $\o_\m{}^{ab}$, $f_\m{}^a$ and $\f_\m^\a$ are expressed as composites of the independent fields $e_\m{}^a$, $A_\m$ and $\j_{\m}$ by imposing suitable curvature constraints, while $b_\m$ is set to zero by gauge-fixing local the special conformal symmetry.}
\label{gauging}
\end{table}

Gauge-fixing special conformal transformations, the local symmetries of $\cn=1$ conformal supergravity comprise diffeomorphisms, Weyl transformations, local frame rotations, $U(1)$ $R$-symmetry, and Q- and S-supersymmetry transformations. The corresponding local parameters, collectively denoted by $\vth$, are shown in table \ref{gauging}. Under these, the independent fields of conformal supergravity transform as
\bal\label{sugra-transformations}
\d(\vth) e_\m{}^a=&\;\cl_\x e_\m{}^a-\l^a{}_b e_\m{}^b+\s e_\m{}^a-\tfrac12\lbar\j_\m\g^a\ve\,,\NO\\
\d(\vth)\j_\m=&\;\cl_\x\j_\m-\tfrac14\l_{ab}\g^{ab}\j_\m+\tfrac12\s\j_\m - i\th\g^5\j_\m+\dc D_\m\ve-\g_\m\h\,,\NO\\
\d(\vth)A_\m=&\;\cl_\x A_\m+\pa_\m\th+\tfrac{3i}{4}\lbar\f_\m\g^5\ve-\tfrac{3i}{4}\lbar\j_\m\g^5\h\,,
\eal
where $\cl_\x$ is the Lie derivative along the vector field $\x^\m$ and the covariant derivative on the spinor parameter $\ve$ is given by $\dc D_\m\ve=(\pa_\m+\tfrac14\o_\m{}^{ab}(e,\j)\g_{ab}+i\g^5A_\m)\ve=(\nabla_\m+i\g^5A_\m)\ve$.

Assigning the local parameters the transformations
\bal\label{parameter-transformations}
&\d(\vth)\x^\m=\x'^\n\pa_\n\x^\m-\x^\n\pa_\n\x'^\m\,,\quad
\d(\vth)\l^a{}_b=\x^\m\pa_\m\l^a{}_b-[\l',\l]^a{}_b\,,\NO\\ &\d(\vth)\s=\x^\m\pa_\m\s\,,\quad
\d(\vth)\th=\x^\m\pa_\m\th\,,\NO\\
&\d(\vth)\ve=\x^\m\pa_\m\ve-\tfrac14\l_{ab}\g^{ab}\ve+\tfrac12\s\ve-i\th\g^5\ve\,,\NO\\
&\d(\vth)\h=\x^\m\pa_\m\h-\tfrac14\l_{ab}\g^{ab}\h-\tfrac12\s\h+i\th\g^5\h\,.
\eal
the local symmetry transformations \eqref{sugra-transformations} obey a closed algebra whose only nonzero commutators are
\bal\label{local_algebra}
[\d_D(\x_1),\d_D(\x_2)]=&\;\d_D(\x_{DD})\,,\quad [\d_L(\l_1),\d_L(\l_2)]=\d_L(\l_{LL})\,,\NO\\
[\d_Q(\ve),\d_W(\s)]=&\;\d_S(\h_{QW})\,,\NO\\ [\d_Q(\ve),\d_S(\h)]=&\;\d_W(\s_{QS})+\d_R(\th_{QS})+\d_S(\h_{QS})+\d_L(\l_{QS})\,,\NO\\
[\d_Q(\ve_1),\d_Q(\ve_2)]=&\;\d_D(\x_{QQ})+\d_L(\l_{QQ})+\d_Q(\ve_{QQ})+\d_R(\th_{QQ})\NO\\
&\;+\d_S(\h_{QQ})\,,
\eal
with composite parameters
\bal\label{composite-parameters}
&\x_{DD}^\m=\x_1^\n\pa_\n\x^\m_2-\x_2^\n\pa_\n\x^\m_1\,,\hspace{.1cm} \l_{LL}{}^a{}_b=[\l_2,\l_1]^a{}_b\,, \hspace{.1cm} \h_{QW}=\tfrac12\g^\m\ve\,\pa_\m\s\,,\NO\\
&\s_{QS}=\tfrac12\lbar\ve\h\,,\hspace{.05cm}\l_{QS}{}^a{}_b=-\tfrac12\lbar\ve\g^a{}_b\h\,,\hspace{.05cm} \th_{QS}=-\tfrac{3i}{4}\lbar\ve\g^5\h\,,\hspace{.05cm} \h_{QS}=\tfrac14(\lbar\h\j_\m)\g^\m\ve\,,\NO\\
& \ve_{QQ}=-\x_{QQ}^\m\,\j_\m\,,\hspace{.1cm} \h_{QQ}=-\x_{QQ}^\m\f_\m+\tfrac14((\lbar\f_\m\ve_1)\g^\m\ve_2-(\lbar\f_\m\ve_2)\g^\m\ve_1)\,,\NO\\
&\x_{QQ}^\m=\tfrac12\lbar\ve_2\g^\m\ve_1\,,\hspace{.1cm} \l_{QQ}^a{}_b=-\x_{QQ}^\m\,\o_\m{}^a{}_b\,,\hspace{.1cm} \th_{QQ}=-\x_{QQ}^\m\,A_\m\,.
\eal

In a scheme where the mixed gravitational-U(1)$_R$ anomaly is placed entirely in the R-symmetry, the partition function transforms under the conformal supergravity symmetries as  
\be\label{W-trans}
\d(\vth)\cw[e,A,\psi]=\int d^4x\, e\,\big(\s\mathcal A_W-\theta\mathcal A_R-\bar\ve\mathcal A_Q+\bar\h\mathcal A_S\big)\,,
\ee
where the anomalies $\ca_W$, $\ca_R$, $\ca_Q$ and $\ca_S$ are local functionals of the independent supergravity fields $e_\m{}^a$, $A_\m$ and $\j_\m$. The Weyl and R-symmetry anomalies were first obtained in \cite{Anselmi:1997am} and were later corrected in \cite{Cassani:2013dba}. The Q- and S-supersymmetry anomalies were first computed holographically for the case of equal $a$ and $c$ anomaly coefficients and to leading order in the gravitino in \cite{Papadimitriou:2017kzw}. They were later determined for arbitrary $a$ and $c$ by solving the WZ consistency conditions for conformal supergravity in \cite{Papadimitriou:2019gel}, and more recently to all orders in the gravitino using BRST techniques in \cite{Imbimbo:2023sph,Imbimbo:2025ffw}. In the accompanying article \cite{suzywzactions}, we solve again the WZ consistency conditions for the conformal supergravity algebra \eqref{local_algebra} and determine the form of all the anomalies $\ca_W$, $\ca_R$, $\ca_Q$ and $\ca_S$ in a two-parameter family of schemes, again to leading nontrivial order in the gravitino. This new result corrects the form of the $\ca_Q$ anomaly obtained in \cite{Papadimitriou:2019gel} for the case $a\neq c$ and provides the explicit form of the $\co(\j^2)$ terms of $\ca_W$, $\ca_R$ that were not given in \cite{Papadimitriou:2019gel}.

\section{Ferrara-Zumino current multiplet and anomalies}

The Ferrara-Zumino (FZ) current multiplet is defined by coupling an SQFT -- not necessarily an SCFT -- to old-minimal supergravity \cite{Komargodski:2010rb}. The field content of old-minimal supergravity comprises the vielbein $\hat e_\m{}^a$, a vector field $\hat A_\m$, a gravitino $\hat\j_\m$ and a complex scalar $\hat\cf$ \cite{Stelle:1978ye,Ferrara:1978em}. The partition function $\cw_{\text{om}}[\hat e,\hat A,\hat\j,\hat\cf]$ defines the FZ current multiplet of operators 
\bal\label{consistent-om-currents}
&\<\hat\ct^\m{}_a\>=\hat e^{-1}\frac{\d\mathcal W_{\text{om}}}{\d \hat e_\m{}^a}\,,\quad \<\hat\cj^\m\>=\hat e^{-1}\frac{\d\mathcal W_{\text{om}}}{\d \hat A_\m}\,,\NO\\
&\<\hat\cs^\m\>=\hat e^{-1}\frac{\d\mathcal W_{\text{om}}}{\d \lbar{\hat\psi}_\m}\,,\quad \<\hat\co_{\cf}\>=\hat e^{-1}\frac{\d\mathcal W_{\text{om}}}{\d \hat\cf}\,.
\eal

For a generic SQFT, all operators in the FZ multiplet are a priori non-local, and the local symmetries acting on the old-minimal supergravity fields are diffeomorphisms, local Lorentz transformations and Poincar\'e supersymmetry, namely
\bal\label{eq:OM-transformations}
\hat\d(\hat\vth_{\text{SP}}) \hat e_\m{}^a=&\;\cl_\x\hat e_\m{}^a-\l^a{}_b \hat e_\m{}^b+\tfrac12\bar\hve \g^a\hat\j_\m\,,\NO\\
\hat\d(\hat\vth_{\text{SP}})\hat A_\m=&\;\cl_\x \hat A_\m+\tfrac{3i}{4}\bar\hve \g^5(\hat\f_\m-\hat\h_0(\hat\j_\m))+\tfrac12\bar\hve \hat\g^\n\hat\j_\m \hat A_\n\,,\NO\\
\hat\d(\hat\vth_{\text{SP}})\hat\j_\m=&\;\cl_\x\hat \j_\m-\tfrac14\l_{ab}\g^{ab}\hat\j_\m +\hat\cd_\m\hve-\hat\g_\m\hat\h_0(\hat\ve)\,,\NO\\
\hat\d(\hat\vth_{\text{SP}}) \hat \cf=&\;\cl_\x\hat\cf-\bar\hve_R\hat\g^\m(\hat\f_\m-\hat\h_0(\hat\j_\m))\,,
\eal
where
\be\label{eq:hateta0}
\hat\h_0(\hve)\equiv\tfrac{i}{3}\hat A_\n\hat\g^\n\g^5\hve-\tfrac12(\hat\cf P_L+\bar{\hat\cf}P_R)\hve\,,
\ee
and all composite quantities are those of conformal supergravity but evaluated on the old-minimal fields. These transformations satisfy the super-Poincar\'e algebra
\bal\label{eq:super-Poincare-algebra}
&[\d_D(\x_1),\d_D(\x_2)]=\d_D(\x_{DD})\,,\quad [\d_L(\l_1),\d_L(\l_2)]=\d_L(\l_{LL})\,,\NO\\
&[\hat \d_Q(\hve_1),\hat \d_Q(\hve_2)]=\d_D(\hat\x_{QQ})+\d_L(\hat\l_{QQ})+\hat\d_Q(\hve_{QQ})\,,
\eal
with field-dependent parameters
\bal\label{eq:super-Poincare-parameters}
&\x_{DD}^\m=\x_1^\n\pa_\n\x^\m_2-\x_2^\n\pa_\n\x^\m_1\,,\quad \l_{LL}{}^a{}_b=[\l_2,\l_1]^a{}_b\,,\NO\\
&\hat\x_{QQ}^\m=\tfrac12\lbar\hve_2\hat\g^\m\hve_1\,,\quad \hve_{QQ}=-\hat\x_{QQ}^\m\,\hat\j_\m\,,\NO\\ &\hat\l_{QQ}^a{}_b=-\hat\x_{QQ}^\m\,\hat\o_\m{}^a{}_b-\tfrac12\lbar\hve_1\g^a{}_b\hat\h_0(\hve_2)+\tfrac12\lbar\hve_2\g^a{}_b\hat\h_0(\hve_1)\,.
\eal

The number of off-shell degrees of freedom encoded in the old-minimal supergravity fields up to the super-Poincar\'e symmetries \eqref{eq:OM-transformations} are 12+12: 6 in $\hat e_\m{}^a$, 4 in $\hat A_\m$, 2 in $\hat\cf$ and 12 in $\hat\j_\m$. However, when an SCFT is coupled to old-minimal supergravity, instead of a generic SQFT, the counting of degrees of freedom is different due to the presence of additional local symmetries. In particular, as in the superspace formulation of $\cn=1$ supergravity \cite{Howe:1978km,Buchbinder:1998twe,Kuzenko:2019vvi}, there exists a chiral multiplet of super-Weyl symmetries, with local parameters $\hat\vth_{\text{SW}}=(\s,\th,\hat\h,\hat\L)$, acting on the old-minimal fields as \cite{suzywzactions}
\bal\label{eq:OM-super-Weyl-transformations}
\hat\d(\hat\vth_{\text{SW}}) \hat e_\m{}^a=&\;\s \hat e_\m{}^a\,,\NO\\
\hat\d(\hat\vth_{\text{SW}})\hat A_\m=&\;\pa_\m\th-\tfrac{3i}{4}\bar{\hat{\h}}\g^5\hat\j_\m\,,\NO\\
\hat\d(\hat\vth_{\text{SW}})\hat\j_\m=&\;\tfrac12\s\hat\j_\m - i\th\g^5\hat\j_\m-\hat\g_\m\hat\h\,,\NO\\
\hat\d(\hat\vth_{\text{SW}}) \hat \cf=&\;-\s \hat\cf+2i\th \hat\cf+\hat{\L}\,.
\eal
When counted up to the extended super-Poincar\'e and super-Weyl symmetries, the off-shell degrees of freedom encoded in the old-minimal fields is 8+8, exactly as in conformal supergravity. In particular, $\hat\cf$ becomes pure gauge and the corresponding FZ multiplet operator, $\<\hat\co_{\cf}\>$, is {\em ultralocal} -- a polynomial in the old-minimal fields and their derivatives.    

Since the super-Weyl symmetries are all Abelian, the only nonzero commutation relations of the extended algebra beyond the super-Poincar\'e ones in \eqref{eq:super-Poincare-algebra} are those between Poincar\'e supersymmetry and the super-Weyl symmetries. These take the form 
\bal\label{eq:OM-algebra-mixed}
[\hat\d_Q(\hve),\hat\d_{W}(\s)]=&\;\hat\d_{S}(\hat{\h}_{QW})+\hat\d_{F}(\hat{\L}_{QW})\,,\NO\\
[\hat\d_Q(\hve),\hat\d_{R}(\th)]=&\;\hat\d_{S}(\hat{\h}_{QR})+\hat\d_{F}(\hat{\L}_{QR})\,,\NO\\ 
[\hat\d_Q(\hve),\hat\d_{S}(\hat{\h})]=&\;\hat\d_{W}(\s_{QS})+\hat\d_{R}(\th_{QS})+\hat\d_L(\l_{QS})+\hat\d_{S}(\hat{\h}_{QS})+\hat\d_{F}(\hat{\L}_{QS})\,,\NO\\
[\hat\d_Q(\hve),\hat\d_{F}(\hat{\L})]=&\;\hat\d_{S}(\hat{\h}_{QF})+\hat\d_{F}(\hat{\L}_{QF})\,,
\eal
with composite parameters
\bal
\hat{\h}_{QW}=&\; \tfrac12\hat\g^\n\hve\,\pa_\n\s\,,\quad \hat{\L}_{QW}=\tfrac12\bar\hve_R\hat\g^\m\hat\g^\n\hat\j_\m\pa_\n\s\,,\NO\\
\hat{\h}_{QR}=&\;\tfrac{i}{3}\g^5\g^\n\hve\,\pa_\n\th\,,\quad \hat{\L}_{QR}=\tfrac{i}{3}\bar{\hve}_R\hat\g^\m\hat\g^\n\hat\j_\m\pa_\n\th\,,\NO\\
\s_{QS}=&\;\tfrac12\bar\hve\hat{\h}\,,\quad \th_{QS}=-\tfrac{3i}{4}\bar\hve\g^5\hat{\h}\,,\quad \l^{ab}_{QS}=-\tfrac12\bar\hve\g^{ab}\hat{\h}\,,\NO\\
\hat{\h}_{QS}=&\;\tfrac14(\bar{\hat\h}\hat\j_\n)\hat\g^\n\hve-\tfrac14(\bar{\hat\h}\g^5\hat\j_\n)\hat\g^\n\g^5\hve\,,\NO\\
\hat{\L}_{QS}=&\;\bar\hve_R\hat\g^\m(\hat\cd_\m\hat\h+\tfrac{2i}{3}\hat A_\m\hat\h+\tfrac12(\bar{\hat\j}_{\n L}\hat\h)\hat\g^\n\hat\j_\m)-\hat\cf\bar\hve_L\hat\eta\,,\NO\\
\hat{\h}_{QF}=&\;\tfrac12(\hat{\L} P_L+\bar{\hat{\L}} P_R)\hve\,,\quad \hat{\L}_{QF}=\tfrac12\hat{\L}\,\bar{\hve}\hat\g^\m\hat\j_{\m\,L}\,.
\eal

The extended algebra \eqref{eq:super-Poincare-algebra}, \eqref{eq:OM-algebra-mixed} determines the old-minimal superconformal anomalies through the corresponding WZ consistency conditions. In contrast to conformal supergravity, in this case there exists a scheme where Poincar\'e supersymmetry in non-anomalous. In this specific scheme, and placing again the mixed gravitational-U(1)$_R$ anomaly entirely in the super-Weyl R-symmetry, a generic symmetry transformation of the old-minimal partition function takes the form
\bal\label{eq:sWanomalies}
&\hat\d(\hat\vth)\cw_{\text{om}}[\hat e,\hat A,\hat\j,\hat\cf]\NO\\
&\hspace{.4cm}=\int d^4x\,\hat e\,\big(\s\ca_W^{\text{om}}-\th\ca_R^{\text{om}}+\bar{\hat\h}\ca_S^{\text{om}}+\hat\L\ca_F^{\text{om}}+\bar{\hat\L}\ca_{\bar F}^{\text{om}}\big)\,,
\eal
where the anomalies $\ca_W^{\text{om}}$, $\ca_R^{\text{om}}$, $\ca_S^{\text{om}}$ and $\ca_F^{\text{om}}$ are local functionals of the old-minimal fields $\hat e_\m{}^a$, $\hat A_\m$, $\hat\j_\m$ and $\hat\cf$. They are the components of the superspace superconformal anomaly and they are determined explicitly in \cite{suzywzactions}.

Notice that \eqref{eq:sWanomalies} determines completely the form of the scalar operator $\<\hat\co_{\cf}\>$, forcing it to be ultralocal. Since super-Weyl $\hat\L$-transformations act on $\hat\cf$ by an additive shift, it follows that 
\bal\label{eq:ulatralocalOF}
\<\hat\co_{\cf}\>=&\;\ca_F^{\text{om}}\NO\\
=&\;-\tfrac{a}{8\p^2}\Big(\big(2\hat\cf\bar{\hat\cf}+\tfrac{1}{3}\hat R+\tfrac{40}{9}\hat A^2-\tfrac{1}{3}\bar{\hat\j}_\m\hat\g^{\m\n}\hat\f_\n\big)\bar{\hat\cf}\NO\\
&\;-2\lbar{\hat\F}^\m\hat\F_{\m R}-\tfrac{4i}{3}(\pa_\m\hat A^\m\bar{\hat\cf}+2\hat A^\m\pa_\m\bar{\hat\cf})\NO\\
&\;+\tfrac{2i}{3}\bar{\hat\cf}\hat A^\m\bar{\hat\j}_\m\hat\g^\n\hat\j_\n-2\bar{\hat\j}^\m\bar{\hat\cf}\hat\F_{\m L}-\tfrac{8i}{3}\hat A^\m\bar{\hat\j}_{\m L}\hat\g^\n\hat\F_\n\Big)\NO\\
&\;+\tfrac{(a-c)}{16\p^2}\lbar{\hat{\mathscr{R}}}_{\m\n}(Q)\hat{\mathscr{R}}(Q)^{\m\n}_L\,,
\eal
where $\hat{\mathscr{R}}(Q)_{\m\n}$ is the gravitino curvature in \eqref{curvatures} evaluated on the old-minimal fields and $\hat\F_\m\equiv\hat\f_\m-\hat\eta_0(\hat\j_\m)$. In particular, for any SCFT coupled to old-minimal supergravity in a scheme where Poincar\'e supersymmetry in non-anomalous, the FZ multiplet operator $\<\hat\co_{\cf}\>$ is universally given by \eqref{eq:ulatralocalOF} -- up to possible contributions from dimension-four supersymmetric invariants that parameterize different supersymmetric schemes \cite{Ferrara:1978rk,Ferrara:1988qx}.

\section{Old-minimal from conformal supergravity}

The descriptions of conformal and old-minimal supergravities we presented so far are a priori independent. However, it is well known that old-minimal supergravity can be obtained from conformal supergravity using conformal calculus \cite{Kaku:1978ea}.

Under the local symmetries of conformal supergravity, a chiral multiplet $(\t,\c,\vf,\cf)$ of Weyl weight 1 transforms as\footnote{See \cite{suzywzactions} for the field redefinitions that relate the fields $(\t,\c,\vf,\cf)$ to the components of a canonical chiral multiplet.}
\bal\label{eq:CompensatorMultipletTransformations}
\d(\vth)\t=&\;\cl_\x\t+\s-\tfrac12\bar\ve\,\vf\,,\NO\\
\d(\vth)\c=&\;\cl_\x\c+\th+\tfrac{3i}{4}\bar\ve\g^5\vf\,,\NO\\
\d(\vth)\vf_L=&\;\cl_\x\vf_L-\tfrac{1}{4}\l_{ab}\g^{ab}\vf_L-\tfrac12 \s\vf_L+i\th\vf_L\NO\\
&-\tfrac{1}{2}(\cd_\n\t+\tfrac{2i}{3}\cd_\n\c+\bar\psi_{\n L}\vf_L)\g^\n\ve_R+\tfrac12(\cf+\bar\vf_L\vf_L)\ve_L+\h_L\,,\NO\\
\d(\vth)\cf=&\;\cl_\x\cf-\s \cf+2i\th \cf+(2\bar\ve_R\vf_R-\bar\ve_L\vf_L)\cf\NO\\
&-\bar\ve_R\g^\m\big(\phi_{\m}-\cd_\m\vf-(\bar\vf_L\j_{\m L})\vf+\tfrac12\cf(\j_{\m}+\g_\m\vf)\NO\\
&-\tfrac12(\cd_\n \t+\tfrac{2i}{3}\cd_\n\c+\bar\j_{\n L}\vf_L)\g^\n(\j_{\m}+\g_\m\vf)\big)\,,
\eal
where the covariant derivatives are given by
\be
\cd_\m\t=\pa_\m\t\,,\quad \cd_\m\c=\pa_\m\c-A_\m\,,\quad \cd_\m\vf=\nabla_\m\vf-i \g^5 A_\m\vf\,.
\ee

Since the fields $\t$, $\c$ and $\vf$ transform by an additive shift under respectively Weyl, R-symmetry and S-supersymmetry, they act as {\em compensators} for the corresponding symmetry. This enables us to construct composite fields that are invariant under these symmetries, namely
\bal\label{eq:InvariantFields}
\hat e_\m{}^a =&\;e^{-\t}\,e_\m{}^a\,,\NO\\
\hat A_\m =&\; A_\m-\del_\m\c+\tfrac{3i}{4}\, \bar\psi_{\m}\g^5\,\vf-\tfrac{3i}{8}\,\bar\vf\g_\m\g^5\vf\,,\NO\\
\hat\psi_{\m } =&\;U\,(\psi_{\m }+\,\g_\m \vf)\,,\NO\\
\hat\cf =&\;e^{\t-2i\c}\cf\,,
\eal
where
\be\label{eq:U}
U\equiv e^{-\tfrac{\t}{2}+ i\g^5\c}=e^{-\tfrac{\t}{2}}(\cos \c+ i\g^5\sin \c)\,.
\ee
Moreover, Q-supersymmetry, together with diffeomorphisms and local Lorentz transformations, act on the composite fields \eqref{eq:InvariantFields} exactly as local super-Poincar\'e symmetry acts on the old-minimal fields in \eqref{eq:OM-transformations} \cite{suzywzactions}. Hence, the fields of old-minimal supergravity can be identified with the composites \eqref{eq:InvariantFields}  \cite{Kaku:1978ea}.

The canonical construction of old-minimal supergravity via the procedure described above involves gauge-fixing Weyl, R-symmetry and S-supersymmetry by setting the compensator fields $\t$, $\c$ and $\vf$ to zero. However, super-Weyl transformations act non-trivially on these fields, and so setting them to zero obscures the role of super-Weyl symmetry. In \cite{suzywzactions} we identify the local super-Poincar\'e algebra \eqref{eq:super-Poincare-algebra} as a subalgebra of the conformal supergravity algebra \eqref{local_algebra} on the entire field space spanned by the conformal supergravity fields and the fields in the compensator multiplet {\em without} any gauge-fixing. From the semi-direct product of the conformal supergravity symmetries with super-Weyl transformations we are then able to construct an old-minimal version of super-Weyl transformations that acts on the invariant fields \eqref{eq:InvariantFields} exactly as in \eqref{eq:OM-super-Weyl-transformations}.

\section{The interpolating local counterterm}

We have argued that any renormalized SCFT can be coupled to either conformal or old-minimal supergravity, defining respectively the conformal and FZ current multiplets of the theory. Coupling the theory to old-minimal supergravity gives the partition function $\cw_{\text{om}}[\hat e,\hat A,\hat\j,\hat\cf]$. Coupling to conformal supergravity instead leads to the partition function $\cw[e,A,\j]$ in \eqref{sugraW}. In this case we can perform finite Weyl, R-symmetry and S-supersymmetry transformations on the conformal supergravity fields using the compensators $\t$, $\c$ and $\vf$ as parameters so that the new background fields are the composites \eqref{eq:InvariantFields}. This results in the partition function $\cw[\hat e,\hat A,\hat\j]$. However, this is not equal to the original partition function $\cw[e,A,\j]$; the difference is given by the WZ effective action 
\be\label{WZdiff}
\cw_{\text{WZ}}[e,A,\j,\t,\c,\vf]=-(\cw[\hat e,\hat A,\hat\j]-\cw[e,A,\j])\,,
\ee
which we compute explicitly in \cite{suzywzactions}. 

The WZ action lifts the conformal supergravity partition function to a functional of the old-minimal fields, namely
\be
\cw[\hat e,\hat A,\hat\j]=\cw[e,A,\j]-\cw_{\text{WZ}}[e,A,\j,\t,\c,\vf]\,.
\ee
It is natural to ask what is the relation between this functional and the old-minimal partition function $\cw_{\text{om}}[\hat e,\hat A,\hat\j,\hat\cf]$. Notice that the transformation of these functionals under the extended symmetries of old-minimal supergravity are different. Most crucially, $\cw_{\text{om}}[\hat e,\hat A,\hat\j,\hat\cf]$ is invariant under Poincar\'e supersymmetry, but $\cw[\hat e,\hat A,\hat\j]$ is not; it transforms as 
\be\label{eq:PoincareQAnomaly1}
\hspace{-.2cm}\hat\d_Q(\hve)\cw[\hat e,\hat A,\hat\j]=\hspace{-.1cm}\int d^4x\hat e(-\bar\hve\ca_Q[\hat e,\hat A,\hat\j]+\bar{\hat\h}_0(\hve)\ca_S[\hat e,\hat A,\hat\j])\,,
\ee
where $\ca_Q[\hat e,\hat A,\hat\j]$, $\ca_S[\hat e,\hat A,\hat\j]$ are the corresponding conformal supergravity anomalies evaluated on the old-minimal fields. Conversely, $\cw[\hat e,\hat A,\hat\j]$ is invariant under super-Weyl $\hat\L$-transformations, while $\cw_{\text{om}}[\hat e,\hat A,\hat\j,\hat\cf]$ is not.

In  \cite{suzywzactions} we show that there exist a local counterterm that depends on the old-minimal fields, $\cw_{\text{ct}}[\hat e,\hat A,\hat\j,\hat\cf]$, such that
\be\label{eq:OMW}
\cw_{\text{om}}[\hat e,\hat A,\hat\j,\hat\cf]=\cw[\hat e,\hat A,\hat\j]+\cw_{\text{ct}}[\hat e,\hat A,\hat\j,\hat\cf]\,.
\ee
The local counterterm splits naturally into two parts: one that depends on the field $\hat\cf$ and one that does not. In particular,
\be\label{eq:OM-counterterm-sum}
\cw_{\text{ct}}[\hat e,\hat A,\hat\j,\hat\cf]=\cw_{\text{ct}}^{(\hat\cf)}[\hat e,\hat A,\hat\j,\hat\cf]+\cw_{\text{ct}}^{(0,0)}[\hat e,\hat A,\hat\j]\,,
\ee
where the $\hat\cf$-dependent part is given by
\bal\label{eq:OM-counterterm-F}
&\;\cw_{\text{ct}}^{(\hat\cf)}[\hat e,\hat A,\hat\j,\hat\cf]=\NO\\
&\;-\tfrac{a}{8\p^2}\int d^4x\,\hat e\,\Big(\hat\cf^2\bar{\hat\cf}^2-2(\bar{\hat\cf}\lbar{\hat\F}^\m\hat\F_{\m L}+\hat\cf\lbar{\hat\F}^\m\hat\F_{\m R})+\tfrac{1}{3}\hat\cf\bar{\hat\cf}\hat R\NO\\
&+\tfrac{40}{9}\hat\cf\bar{\hat\cf}\hat A^\m\hat A_\m+\tfrac{4i}{3}\hat A^\m(\bar{\hat\cf}\pa_\m\hat\cf-\hat\cf\pa_\m\bar{\hat\cf})-\tfrac{1}{3}\hat\cf\bar{\hat\cf}\bar{\hat\j}_\m\hat\g^{\m\n}\hat\f_\n\NO\\
&+\tfrac{2i}{3}\hat\cf\bar{\hat\cf}\hat A^\m\bar{\hat\j}_\m\g^5\hat\g^\n\hat\j_\n+\tfrac{8i}{3}\hat A^\m(\bar{\hat\cf}\bar{\hat\j}_{\m R}\hat\g^\n\hat\F_\n-\hat\cf\bar{\hat\j}_{\m L}\hat\g^\n\hat\F_\n)\Big)\NO\\
&+\tfrac{(a-c)}{16\p^2}\int d^4x\,\hat e\,\big(\hat\cf\lbar{\hat{\mathscr{R}}}_{\m\n}(Q)\hat{\mathscr{R}}(Q)^{\m\n}_L+\bar{\hat\cf}\lbar{\hat{\mathscr{R}}}_{\m\n}(Q)\hat{\mathscr{R}}(Q)^{\m\n}_R\big)\,,
\eal
and the $\hat\cf$-independent part takes the form
\bal\label{eq:OM-counterterm-0}
&\;\cw^{(0,0)}_{\rm ct}[\hat e,\hat A,\hat\j]=\NO\\
&\;-\tfrac{a}{\p^2}\int d^4x\,\hat e\,\big(-\tfrac49\big(\hat f^{\m\n}-\tfrac1{18}\hat A^\m\hat A^\n+\tfrac{1}{8}\bar{\hat\j}_\r\hat\g^\m\hat\g^{\r\s}\hat\g^\n\hat\f_{\s}\big)\hat A_\m \hat A_\n\NO\\
&\hspace{.5cm}+\tfrac{i}{6}\bar{\hat\F}_{0 \m}(2\hat A^\m\hat\g^\n-\hat A_\r\hat\g^\r\hat\g^{\m\n})\g^5\hat\F_{0 \n}\big)\NO\\
&\;+\tfrac{(a-c)}{72\p^2}\int d^4x\,\hat e\,\hat\e^{\m\n\r\s}\hat A_\m\big(i \hat F_{\r\s}\bar{\hat\j}^\k\g^5(2\hat\g_\k\hat\j_\n+\hat\g_\n\hat\j_\k)\NO\\
&\hspace{.5cm}-\tfrac12\hat W_{\r\s\k\l}\bar{\hat\j}^\k(2\hat\g^\l\hat\j_\n-\hat\g_\n\hat\j^\l)+\co(\j^4)\big)\,.
\eal
As above, $\hat{\mathscr{R}}(Q)_{\m\n}$ is the gravitino curvature in \eqref{curvatures} evaluated on the old-minimal fields and we have defined $\hat\F_\m\equiv\hat\f_\m-\hat\eta_0(\hat\j_\m)$ and 
$\hat\F_{0 \m}\equiv\hat\f_\m-\hat\eta_0(\hat\j_\m)\big|_{\hat\cf=\bar{\hat\cf}=0}$. The expression \eqref{eq:OM-counterterm-F} for the $\hat\cf$-dependent part of the local counterterm is exact in the gravitino. Its derivative with respect to $\hat\cf$ determines the ultralocal form of the FZ multiplet operator $\<\hat\co_{\cf}\>$ in \eqref{eq:ulatralocalOF}. The $\hat\cf$-independent part in \eqref{eq:OM-counterterm-0} is correct to up to $\co(\j^4)$ terms and is given here for the specific values  
$s_1=0$ and $s_2=1$ of the scheme parameters introduced in \cite{suzywzactions}.

As we demonstrate in \cite{suzywzactions}, the superspace counterterm constructed in \cite{Kuzenko:2019vvi} coincides with the $\hat\cf$-dependent part \eqref{eq:OM-counterterm-F}. However, both the $\hat\cf$-dependent and $\hat\cf$-independent parts of the local counterterm \eqref{eq:OM-counterterm-sum} are required in order to interpolate between old-minimal and conformal supergravity.

\section{Discussion}

We have determined the local counterterm that interpolates between the conformal and Ferrara-Zumino multiplets of generic $\cn=1$ SCFTs in four dimensions, defined respectively by $\cn=1$ conformal and old-minimal supergravity. The complete counterterm is given in components in \eqref{eq:OM-counterterm-sum} and consists of both an $\hat\cf$-dependent part, given in \eqref{eq:OM-counterterm-F}, and an $\hat\cf$-independent part, given in \eqref{eq:OM-counterterm-0}. The $\hat\cf$-dependent part \eqref{eq:OM-counterterm-F} provides the component expression for the superspace counterterm of \cite{Kuzenko:2019vvi}. However, the $\hat\cf$-independent part in \eqref{eq:OM-counterterm-0} is necessary in order to interpolate between the two multiplets. In particular, the $\hat\cf$-independent part of the counterterm is required in order to interpolate between the standard superspace and conformal supergravity Weyl and R-symmetry anomalies. 

The $\hat\cf$-independent part of the counterterm in \eqref{eq:OM-counterterm-0} is crucial for the comparison of physical observables, such as the supersymmetric Casimir energy \cite{Assel:2015nca,Bobev:2015kza,Martelli:2015kuk,BenettiGenolini:2016qwm,Brunner:2016nyk,Panopoulos:2023cdg} and partition functions \cite{Closset:2013vra,Closset:2014uda,Genolini:2016ecx,Papadimitriou:2017kzw}, computed in the two multiplets. The old-minimal auxiliary field vanishes in common supersymmetric backgrounds of conformal and old-minimal supergravities, and so the $\hat\cf$-dependent part of the counterterm \eqref{eq:OM-counterterm-F} does not play any role in the comparison of most physical observables.

Besides facilitating the comparison of physical observables between the conformal and FZ multiplets of four-dimensional $\cn=1$ SCFTs, our analysis raises the intriguing possibility of systematically constructing certain current multiplets in contexts where they do not arise naturally. For example, off-shell conformal supergravity arises at the conformal boundary of gauged supergravities that admit Anti de Sitter (AdS) vacua. Our embedding of the local super-Poincar\'e algebra within the conformal supergravity algebra without gauge-fixing the compensator multiplet suggests a concrete pathway to the construction of gauged supergravities that produce old-minimal -- rather than conformal -- supergravity on the AdS conformal boundary. The same approach might be applicable to other off-shell multiplets as well, in four or other dimensions.

\clearpage
\begin{acknowledgments}
We would like to thank Piljin Yi and Matthew M. Roberts for valuable discussions. The work of PP is supported by KIAS Grant PG 105601. Part of this work has been supported by the National Research Foundation of Korea (NRF) grant funded by the Korea government (MSIT) (No.~2023R1A2C1006975) and from the JRG Program at the Asia Pacific Center for Theoretical Physics (APCTP) through the Science and Technology Promotion Fund and Lottery Fund of the Korean Government. IP was supported in part by the Brain Pool program funded by the Ministry of Science and ICT through the National Research Foundation of Korea (RS-2025-25457100). PP thanks the National and Kapodistrian University of Athens for  warm hospitality at several stages of this work and the APCTP, where this project was formally initiated. IP thanks the APCTP and Kyung Hee University for the hospitality during the completion of this work. We would also like to acknowledge extensive use of the symbolic computation package FieldsX \cite{Frob:2020gdh}.
\end{acknowledgments}

\bibliography{refs.bib}
\end{document}